\documentclass[acmsmall]{acmart}

\usepackage{enumitem}
\usepackage{amsmath}
\usepackage{mathtools}
\usepackage{bm}
\usepackage{ifthen}

\usepackage{amssymb}

\setcopyright{none}
\renewcommand\footnotetextcopyrightpermission[1]{}

\newtheorem{definition}{Definition}
\newtheorem{condition}{Condition}
\newtheorem{claim}{Claim}

\newtheorem{theorem}{Theorem}

\newboolean{arxiv}

\newcommand{\refdef}[1]{Definition \ref{#1}}
\newcommand{\refcon}[1]{Condition \ref{#1}}
\newcommand{\refclm}[1]{Claim \ref{#1}}

\newcommand{\refequ}[1]{(\ref{#1})}

\newcommand\given[1][]{\:#1\vert\:}

\DeclareMathOperator{\RR}{\mathbb{R}}
\DeclareMathOperator{\EE}{\mathbb{E}}

\setboolean{arxiv}{true}
\begin{document}

\title{Beyond Theorems: A Counterexample to Potential Markov Game Criteria}

\author{Fatemeh Fardno}
\affiliation{
  \institution{University of Waterloo}
  \city{Waterloo}
  \country{Canada}}
\email{ffardno@uwaterloo.ca}

\author{Seyed Majid Zahedi}
\affiliation{
  \institution{University of Waterloo}
  \city{Waterloo}
  \country{Canada}}
\email{smzahedi@uwaterloo.ca}

\begin{abstract}
There are only limited classes of multi-player stochastic games in which independent learning is guaranteed to converge to a Nash equilibrium.
Markov potential games are a key example of such classes.
Prior work has outlined sets of sufficient conditions for a stochastic game to qualify as a Markov potential game.
However, these conditions often impose strict limitations on the game's structure and tend to be challenging to verify.
To address these limitations, Mguni et al.~\cite{mguni2021learning} introduce a relaxed notion of Markov potential games and offer an alternative set of necessary conditions for categorizing stochastic games as potential games.
Under these conditions, the authors claim that a deterministic Nash equilibrium can be computed efficiently by solving a dual Markov decision process.
In this paper, we offer evidence refuting this claim by presenting a counterexample.
\end{abstract}

%

\pagestyle{fancy}
\fancyhead{}
\fancyfoot{}

\maketitle
\thispagestyle{empty}

\section{Introduction}
\label{sec:introduction}
In recent years, there has been a growing interest in applying multi-agent reinforcement learning to find strategies that converge to a Nash equilibrium in multi-player stochastic games~\cite{leonardos2021global,foerster2017learning,macua2018learning,hu2003nash,borkar2002reinforcement,guo2021decentralized,perolat2017learning}.
In single-agent environments, many learning algorithms are guaranteed to converge to optimal policies under some mild conditions~\cite{sutton2018reinforcement,watkins1989learning,watkins1992q}.
However, deploying independent single-agent learning algorithms in a multi-agent environment does not guarantee finding Nash equilibrium policies~\cite{foerster2017stabilising}.
The main reason for this is that, from the perspective of a learning agent, the environment undergoes constant changes as other agents concurrently learn, and these environmental changes partly depend on the actions of the learning agent.

Another challenge in deploying multi-agent learning is efficiency.
The computation of a stationary Nash equilibrium for general stochastic games is known to be computationally intractable~\cite{daskalakis2023complexity}.
Consequently, there are no efficient multi-agent learning algorithms for learning stationary Nash equilibrium strategies in general stochastic games.
However, there are specific classes of stochastic games where Nash equilibrium strategies could be computed efficiently.
One notable class is Markov potential games (MPGs), within which multi-agent learning, and specifically independent learning, exhibits promising convergence properties~\cite{leonardos2021global,macua2018learning,fox2022independent}.

An MPG is characterized by the existence of a global function, known as the \emph{potential function}, where a change in an agent's long-term payoff due to a unilateral change in the agent's strategy equates to the change in the potential function.
This unique property facilitates the use of efficient multi-agent learning methods, such as the independent policy gradient, ensuring convergence to a stationary Nash equilibrium strategy~\cite{leonardos2021global}.
However, determining whether a stochastic game qualifies as an MPG by searching for such a potential function is not always straightforward.
Consequently, given the highly desirable properties of MPGs, it becomes essential to identify sufficient conditions for categorizing games as MPGs.

Existing research has established sets of sufficient conditions for identifying a stochastic game as an MPG~\cite{macua2018learning,leonardos2021global}.
However, these conditions often impose stringent restrictions on the game's structure and may not be easy to verify.
To address these limitations, in a recent work~\cite{mguni2021learning}, the authors introduce a relaxed notion of MPGs and offer an alternative set of necessary conditions for categorizing stochastic games as potential games.
The authors claim that meeting these conditions ensures the existence of a deterministic stationary Nash equilibrium in the game, which corresponds to the optimal solution of a single-agent Markov decision process (MDP) constructed based on the original stochastic game.
Essentially, the deterministic stationary Nash equilibrium of the original stochastic game can be efficiently computed by solving its (dual) MDP\@.
This, in turn, guarantees the convergence of multi-agent learning methods to Nash equilibrium strategies.
In this paper, we scrutinize this key claim and present a counterexample to Theorem 1 in~\cite{mguni2021learning}, establishing a case where the theorem does not hold.
\section{Background and Related Work}
\label{sec:background}

In this section, we first briefly introduce stochastic games.
We then provide some background on Markov potential games.
We then discuss sufficient conditions for a stochastic game to be a Markov potential game.

\subsection{Stochastic Games}
\label{subsec:sgs}
We start by defining \emph{Markov decision processes (MDP)} as a tool to study decision making in single-agent environments:
\begin{definition}[\textbf{MDP}]
\label{def:mdp}
 A Markov decision process is a tuple $( S, A, r, p )$.
 $S$ is the state space.
 $A$ is the action space.
 $r: S \times A \mapsto \RR$ is the payoff function.
 And $ p: S \times A \times S \mapsto [0, 1]$ is the transition probability function.
\end{definition}
\noindent
Please note that in this definition, the action space is assumed to be the same across all states.
Relaxing this assumption introduces additional notation, but apart from that, it does not pose any significant difficulty or insight.

For any MDP, a stationary strategy $\pi: S \times A \mapsto [0, 1]$ maps each state-action pair to a probability.
We write $\pi(a \given s)$ to denote the probability of taking action $a$ in state $s$ under strategy $\pi$.
A strategy $\pi$ is called deterministic if for all states $s \in S$, there exists an action $a \in A$ for which $\pi(a \given s) = 1$.

\emph{Stochastic games} (a.k.a. \emph{Markov games}) extend MDPs to multi-agent environments:
\begin{definition}[\textbf{Stochastic game}]
 An $n$-agent stochastic game is a tuple $( N, S, \bm{A}, \bm{r}, p )$%
 \footnote{We use bold font to represent vectors.}.
 $N$ is the set of agents.
 $S$ is the state space.
 $\bm{A} = A_1 \times \dots \times A_n$, where $A_i$ is the action space of agent $i \in N$.
 $\bm{r} = r_1 \times \dots \times r_n$, where $r_i: S \times \bm{A} \mapsto \RR$ is the payoff function for agent $i \in N$.
 And $p: S \times \bm{A} \times S \mapsto [0, 1]$ is the transition probability function.
\end{definition}
\noindent
Note again that the action space of each agent is assumed to be the same across all states.
Similar to \refdef{def:mdp}, this assumption can be easily removed.

We use $\pi_i$ to denote the strategy of agent $i$.
The joint strategy profile of all agents is denoted by $\bm{\pi} = \pi_1 \times \dots \times \pi_n$.
And the joint strategy profile of all agents except agent $i$ is denoted by $\bm{\pi}_{-i} = \pi_1 \times \dots \times \pi_{i-1} \times \pi_{i+1} \times \dots \times \pi_n$%
\footnote{We use the notation $-i$ to indicate all agents except agent $i$.}.

In infinite-horizon stochastic games, the long-term value of state $s$ to agent $i$ for strategy $\bm{\pi}$ is the expected sum of agent $i$'s discounted payoffs:
\begin{equation}
\label{eq:value}
 V_i^{\bm{\pi}}(s) \triangleq \EE_{\bm{\pi}} \left[ \sum_{t=0}^{\infty} \gamma^t r_i(s_t,\bm{a_t}) \given[\Big] s_0 = s \right],
\end{equation}
where $\EE_{\bm{\pi}}[\cdot]$ denotes the expected value of a random variable given that agents follow joint strategy profile $\bm{\pi}$, and $\gamma$ is the discount factor%
\footnote{$\gamma$ determines how much agents discount future payoffs.}.
Agents are considered to be self-interested. 
Each agent's objective is to maximize its own long-term value. 
A \emph{best-response strategy} is a strategy that achieves the highest value for an agent given other agents' strategies~\cite{blume1995statistical}.
\emph{Nash equilibrium} is a joint strategy where each agent's strategy is a best response to others':
\begin{definition}[\textbf{$\bm{\epsilon}$-Nash equilibrium}]
\label{epsNash}
 Let $\epsilon \ge 0$.
 Then in an $n$-agent stochastic game, an $\epsilon$-Nash equilibrium is a strategy profile $\bm{\pi^*} = \pi^*_1, \dots \pi^*_n$ such that:
 \[
  V_i^{\bm{\pi}^*}(s) \geq V_i^{(\pi_i,\bm{\pi}_{-i}^*)}(s) - \epsilon 
 \]
 for all states $s \in S$, all agents $i \in N$, and all strategies $\pi_i \in \Pi_i$, where $\Pi_i$ is the set of all strategies for agent $i$.
\end{definition}
\noindent
When $\epsilon = 0$, we simply call this a Nash equilibrium.
A Nash equilibrium strategy $\bm{\pi^*}$ is stationary if $\pi^*_i$ is stationary for all agents $i \in N$.
All stochastic games have at least one stationary Nash equilibrium~\cite{fink1964equilibrium}.

Computing a stationary $\epsilon$-Nash equilibrium for general stochastic games is computationally intractable~\cite{daskalakis2023complexity}%
\footnote{In fact, the authors show that computing a stationary coarse-correlated equilibrium, which is a more relaxed notion compared to Nash equilibrium, in general stochastic games is computationally intractable.}.
This implies that there are no efficient multi-agent reinforcement learning algorithms for learning stationary Nash equilibrium strategies in general stochastic games.
However, there are specific classes of stochastic games for which Nash equilibrium strategies could be computed efficiently.
A key example is \emph{Markov potential games (MPG)}.

\subsection{Markov Potential Games}\label{subsec:markov-potential-games}
Monderer and Shapley introduce the concept of potential games in the normal form~\cite{monderer1996potential}.
Potential games represent multi-agent coordination, as all agents' payoffs are perfectly aligned with each other via a potential function.
MPGs extend the concept of potential games from normal-form games to stochastic games.
A stochastic game qualifies as an MPG if there exists a global function, called the \emph{potential function}, such that if any agent unilaterally changes their strategy, the change in their long-term value for each state mirrors precisely the change observed in the potential function at that particular state:
\begin{definition}[\textbf{MPG}]
\label{def:mpg}
 A stochastic game is an MPG if there exists a strategy-dependent function $\phi^{\bm{\pi}}: S \mapsto \RR$ for strategies $\bm{\pi} \in \bm{\Pi}$ such that:
 \[
  V_i^{\bm{\pi}}(s) - V_i^{(\pi^\prime_i, \bm{\pi}_{-i})}(s) = \phi^{\bm{\pi}}(s) - \phi^{(\pi^\prime_i, \bm{\pi}_{-i})}(s)
 \]
 for all agents $i \in N$, all states $s \in S$, all strategies $\bm{\pi} = (\pi_i, \bm{\pi}_{-i}) \in \bm{\Pi}$, and all strategies $\pi_i^\prime \in \Pi_i$.
\end{definition}
\noindent
Any MPG has at least one stationary Nash equilibrium strategy profile that is deterministic~\cite{leonardos2021global}.
Furthermore, independent learning converges to an $\epsilon$-Nash equilibrium strategy in MPGs:
\begin{theorem}[\textbf{\cite[Theorem 1.1]{leonardos2021global}}]
 In an $n$-agent MPG, if all agents run independent policy gradient, then for any $\epsilon > 0$, the learning dynamics reaches an $\epsilon$-Nash equilibrium strategy after $O(1/\epsilon^2)$ iterations.
\end{theorem}
\noindent
The main idea behind the aforementioned theorem is that in MPGs, applying \emph{projected gradient ascent (PGA)} on the potential function $\phi$ leads to the emergence of an $\epsilon$-Nash equilibrium.
And the key element in the proof of this theorem is the equality of the derivatives between value functions and the potential function in MPGs.
More recently, in~\cite{fox2022independent}, the authors show that in an MPG, independent \emph{natural policy gradient} also converges to an equilibrium.

\subsection{From Stochastic Games to MPG}\label{subsec:from-sg-to-mpg}
In \refdef{def:mpg}, the condition is fairly strong and difficult to verify in practice for general stochastic games.
Given the MPGs' desiderata, it becomes imperative to delineate the specific types of stochastic games that align with the criteria outlined in \refdef{def:mpg}.
To this end, prior work has provided sets of sufficient conditions~\cite{macua2018learning,leonardos2021global}.
To discuss these conditions, we first need to define the class of \emph{one-shot potential stochastic games (OPSGs)}.
We define a stochastic game as an OPSG if immediate payoffs at any state are captured by a potential game at that state:
\begin{definition}[\textbf{OPSG}]
\label{def:one-shot}
 An $n$-aagent stochastic game is OPSG if there exists a one-shot potential function $\Phi: S \times \bm{A} \mapsto \RR$ such that:
 \[
  r_i(s,\bm{a}) - r_i(s, a^\prime_i, \bm{a}_{-i}) = \Phi(s, \bm{a}) - \Phi(s, a^\prime_i, \bm{a}_{-i})
 \]
 for all $i \in N$, all states $s \in S$, all action profiles $\bm{a} = (a_i, \bm{a}_{-i}) \in \bm{A}$, and all actions $a^\prime_i \in A_i$.
\end{definition}

In~\cite{leonardos2021global}, the authors show that an OPSG is MPG if either of the two following conditions hold: (i) \emph{agent-independent transitions} and (ii) \emph{equality of individual dummy terms}.
(i) holds if the game's transition probability function does not depend on agents' joint action:
\begin{condition}[\textbf{Agent-independent transitions}]
\label{con:indep-tran}
 An OPSG has agent-independent transitions if for all states $s, s^\prime \in S$ and action profiles $\bm{a} \in \bm{A}$:
 \[
  p(s^\prime \given s, \bm{a}) = p(s^\prime \given s).
 \]
\end{condition}
\noindent
And (ii) holds if the dummy terms of each agent's immediate payoffs are equal across all states:
\begin{condition}[\textbf{Equality of individual dummy terms}]
\label{con:equal-dummy}
 An OPSG with one-shot potential function $\Phi$ satisfies the equality of individual dummy terms if for each agent $i \in N$, there exists a function $v_i: S \times \bm{A}_{-i} \mapsto \RR$ such that:
 \[
  r_i(s, a_i,\bm{a}_{-i}) = \Phi(s, a_i,\bm{a}_{-i}) + v^i(s, \bm{a}_{-i}),
 \]
 and
 \[
  \nabla_{\pi_i(s)}\EE\left[\sum_{t=0}^{\infty} \gamma^t v_{i}(s_t, \bm{a}_{-i}^t) \given[\Big] s_0 = s\right] = c_s \bm{1}
 \]
 for all states $s, s^\prime \in S$, $c_s \in \RR$, and $\bm{1} \in \RR^{\vert A_i \vert}$, where $\pi_i(s)$ is the strategy of agent $i$ at state $s$.
\end{condition}  

In~\cite{mguni2021learning}, the authors argue that \refcon{con:indep-tran} and \refcon{con:equal-dummy} impose strong limitations on the structure of one-shot potential stochastic games.
To avoid these limitations, the authors propose an alternative condition:
\begin{condition}[\textbf{State transitivity}]
\label{con:state-trans}
 An OPSG with one-shot potential function $\Phi$ satisfies state transitivity if we have:
 \[
  r_i(s, \bm{a}) - r_i(s^\prime, \bm{a}) = \Phi(s, \bm{a}) - \Phi(s^\prime, \bm{a})
 \]
 for all agents $i \in N$, all states $s, s^\prime \in S$, and all action profiles $\bm{a} \in A$.
\end{condition}
\noindent
State transitivity ensure that the difference in immediate payoffs for changing state is the same for each agent.
The authors then present a theorem that claims the following.
\begin{claim}[\textbf{\cite[Theorem 1]{mguni2021learning}}]
\label{clm:opsp-mpg}
 Let $G \coloneqq (N,S,\bm{A},\bm{r},p)$ be an OPSG with one-shot potential function $\Phi$.
 Suppose that $G$ satisfies \refcon{con:state-trans}.
 Then $G$ has a deterministic stationary Nash equilibrium that corresponds to the optimal solution of the dual MDP defined as $G^\prime \coloneqq (S, \bm{A}, \Phi, p)$.
 That is, the deterministic stationary Nash equilibrium of $G$ can be efficiently computed by solving $G^\prime$.
\end{claim}

\section{Analysis}
\label{sec:counter}
In this section, we provide a counterexample for which \refclm{clm:opsp-mpg} fails to hold.

\subsection{Counterexample}
\label{subsec:counterexample}
Since action space and state space are assumed to be continuous in~\cite{mguni2021learning}, our counterexample includes continuous action and state spaces.
Similarly, since~\cite{mguni2021learning} focuses on infinite-horizon games, our counterexample considers an infinite-horizon game.

Consider game $G$, an infinite-horizon, two-agent stochastic game defined as:
\begin{itemize}[leftmargin=*,noitemsep=*,topsep=1pt]
 \item The set of agents is $N = \{1, 2\}$.
 \item The continuous state space is $S = [0, 1]$. 
 \item The action spaces are $A_1 = A_2 = [0, 1]$.
 \item The payoff functions for any state $s \in S$ and any action profile $a = (a_1, a_2)$ are:
  \[
   r_1(s, a_1, a_2) = s - (s - a_2)^2 - 4/(2 - a_2), 
  \]
  and
  \[
   r_2(s, a_1, a_2) = s - (s - a_2)^2.
  \]
 \item The state transition function only depends on the action of agent $1$ and can be written as: 
  \[
   p(s^\prime \given s, a_1, a_2) = \begin{cases}
    1 &\text{if } s^\prime = a_1, \text{ and} \\
    0 &\text{otherwise}.
   \end{cases}
  \]
\end{itemize}

It can be easily verified the aforementioned game $G$ satisfies all the assumptions in~\cite{mguni2021learning}.
In particular, the payoff functions are bounded, measurable functions in the actions, Lipschitz, and continuously differentiable in the state and actions.

Next, we show that $G$ is an OPSG\@.
We do this by showing that immediate payoffs at any state are captured by a potential game at that state.
Consider the following potential function:
\[
 \Phi(s, a_1, a_2) = s - (s - a_2)^2.
\]
It is easy to see that $G$ is a potential game at each state $s \in S$ with the potential function $\Phi(s, \cdot)$:
\begin{align*}
 \ifthenelse{\boolean{arxiv}}{
  r_1(s, a_1, a_2) &- r_1(s, a_1^\prime, a_2)  = \Phi(s, a_1, a_2) - \Phi(s, a_1^\prime, a_2) = 0,
 }{
  r_1(s, a_1, a_2) &- r_1(s, a_1^\prime, a_2)  = \Phi(s, a_1, a_2) - \Phi(s, a_1^\prime, a_2)
  \\&= 0,
 }
\end{align*}
and
\begin{align*}
 \ifthenelse{\boolean{arxiv}}{
  r_2(s, a_1, a_2) &- r_2(s, a_1, a_2^\prime)  = \Phi(s, a_1, a_2) - \Phi(s, a_1, a_2^\prime) = (s - a_2^\prime)^2 - (s - a_2)^2.
 }{
  r_2(s, a_1, a_2) &- r_2(s, a_1, a_2^\prime)  = \Phi(s, a_1, a_2) - \Phi(s, a_1, a_2^\prime) \\
  &=(s - a_2^\prime)^2 - (s - a_2)^2.
 }
\end{align*}
It is also easy to see that $G$ satisfies \refcon{con:state-trans} for all states $s, s^\prime \in S$ and action profiles $a=(a_1, a_2)$:
\begin{align*}
 \ifthenelse{\boolean{arxiv}}{
  r_1(s, a_1, a_2) &- r_1(s^\prime, a_1, a_2) = \Phi(s, a_1, a_2) - \Phi(s^\prime, a_1, a_2) = (s - s^\prime) - (s - a_2)^2 + (s^\prime - a_2)^2,
 }{
  r_1(s, a_1, a_2) &- r_1(s^\prime, a_1, a_2) = \Phi(s, a_1, a_2) - \Phi(s^\prime, a_1, a_2) \\
  &=(s - s^\prime) - (s - a_2)^2 + (s^\prime - a_2)^2,
 }
\end{align*}
and
\begin{align*}
 \ifthenelse{\boolean{arxiv}}{
  r_2(s, a_1, a_2) &- r_2(s^\prime, a_1, a_2) = \Phi(s, a_1, a_2) - \Phi(s^\prime, a_1, a_2) = (s - s^\prime) - (s - a_2)^2 + (s^\prime - a_2)^2.
 }{
  r_2(s, a_1, a_2) &- r_2(s^\prime, a_1, a_2) = \Phi(s, a_1, a_2) - \Phi(s^\prime, a_1, a_2) \\
  &=(s - s^\prime) - (s - a_2)^2 + (s^\prime - a_2)^2.
 }
\end{align*}

For contradiction, let us assume that \refclm{clm:opsp-mpg} holds for $G$.
Then we can construct $G$'s dual MDP, $G^\prime$, as follows.
The action space is $\bm{A} = A_1\times A_2$.
The action at each state is $\bm{a} = (a_1, a_2) \in \bm{A}$.
And the payoff function is:
\[
 r(s, (a_1, a_2)) = \Phi(s, a_1, a_2) = s - (s - a_2)^2.
\]
Finally, $G^\prime$ has the same transition probability function as $G$.
It can be easily shown that this MDP has the following unique (deterministic) optimal strategy:
\begin{equation}
\label{eq:pi-star}
 \bm{\pi}^*((a_1, a_2) \given s) = \begin{cases}
  1 &\text{if } (a_1, a_2) = (1, s), \text{ and}\\
  0 &\text{otherwise}.
 \end{cases}
\end{equation}
This optimal joint strategy prescribes taking $a_1 = 1$ and $a_2 = s$ in any state $s \in S$.

Next, we show that this joint strategy profile is not a Nash equilibrium of $G$.
To see this, suppose that agent $2$'s strategy is to take $a_2 = s$ in every state $s$.
By fixing agent $2$'s stationary strategy, we can find agent $1$'s best response by constructing an MDP with the immediate payoff function of:
\begin{equation}
\label{eq:r1}
 r_1(s, a_1) = s - 4/(2 - s).
\end{equation}
In this MDP, agent $1$'s action does not directly affect the immediate payoff at each state.
However, agent $1$'s actions affect the long-term payoff by determining the next states through the transition probability function.
Given \refequ{eq:r1}, agent $1$'s long-term payoff is maximized when $s = 0$.
Hence, the unique optimal strategy of agent $1$ is to take $a_1 = 0$ at every state $s$.
This means that $\bm{\pi}^*$ in \refequ{eq:pi-star} does not correspond to $G$'s stationary Nash equilibrium, a contradiction!

We note that a stationary Nash equilibrium of $G$ is for agent $1$ and $2$ to respectively take $a_1 = 0$ and $a_2 = s$ in all states $s \in S$.
Starting from $s = 0$, the average payoff of agent $1$ under this Nash equilibrium is -2, and the average payoff of agent $2$ is 0.

\section{Conclusion}\label{sec:conclusion}
In this paper, we first introduced stochastic games briefly.
We then provided background information on Markov potential games and discussed the sufficient conditions for a stochastic game to be classified as a Markov potential game.
Furthermore, we examined the main claim of~\cite{mguni2021learning} and presented a counterexample to its Theorem 1, demonstrating that the theorem does not always hold.

\bibliographystyle{ACM-Reference-Format}
\bibliography{arXiv}


\end{document}